\documentclass{report}
\usepackage[T1]{fontenc}
\usepackage[utf8]{inputenc}
\usepackage[magyar]{babel}
\addto\captionsmagyar{
  \renewcommand{\contentsname}%
    {Contents}%

}

\usepackage{indentfirst}
\usepackage{amsmath}
\usepackage{amssymb}
\usepackage{graphicx}
\usepackage{epstopdf}
\usepackage{xcolor}
\usepackage{epstopdf}
\usepackage{url}
\usepackage{hyperref}
\usepackage{caption}

\title{Multiple sequence alignment for short sequences} 

\author{{Kristóf Takács\textsuperscript{*}}} 

\date{}

\begin{document}
\newtheorem{thm}{Theorem}

\newtheorem{defi}{Definition}
 \pagenumbering{gobble} 



\maketitle 

\setcounter{tocdepth}{2} 

\tableofcontents 


\section*{Abstract} 

Multiple sequence alignment (MSA) has been one of the most important problems in bioinformatics for more decades and it is still heavily examined by many mathematicians and biologists. However, mostly because of the practical motivation of this problem, the research on this topic is focused on aligning long sequences. It is understandable, since the sequences that need to be aligned (usually DNA or protein sequences) are generally quite long (e. g., at least 30-40 characters). Nevertheless, it is a challenging question that exactly where MSA starts to become a real hard problem (since it is known that MSA is $\mathbf{NP}$-complete \cite{Elias}), and the key to answer this question is to examine short sequences. If the optimal alignment for short sequences could be determined in polynomial time, then these results may help to develop faster or more accurate heuristic algorithms for aligning long sequences. In this work, it is shown that for length-1 sequences using arbitrary metric, as well as for length-2 sequences using unit metric, the optimum of the MSA problem can be achieved by the trivial alignment.



{\let\thefootnote\relax\footnotetext{* \textit{
Eötvös Loránd University, Department of Computer Science, Budapest, Hungary}}}


\newpage 

\clearpage
\pagenumbering{arabic} 
\setcounter{page}{1}


\chapter*{1. Introduction}
\addcontentsline{toc}{chapter}{1. Introduction}

Aligning more than two sequences with as little cost as possible is a quite essential problem for those who are interested in bioinformatical research. E. g., by this method, some conjecture can be given which are the conservative regions of some particular sequences that can generally determine the basic functions and parameters of a group of DNA, RNA or protein sequences \cite{Kumar}. It is also worth to mention that multiple sequence alignment (MSA) is one of the most important tools that is used during motif finding: e. g., if one is given some gene sequences so that is known they perform the same function in different species, it is a plausible question that exactly what these sequences in common have? Among other processes, MSA can be a very useful method to answer questions like this \cite{Frith}.   

Considering the practical significance of multiple sequence alignment, it is not surprising that this problem is in the centre of bioinformatical research for decades. The practical motivation of this problem indicated that most people who attended to MSA have tried to find a new result for the general question, construct a better approximation algorithm or prove something about the complexity of MSA. It was probably the most important achievement in this topic when Isaac Elias has proved that MSA is $\mathbf{NP}$-complete if the score scheme of the characters is a metric \cite{Elias}. (This negative result can be even more interesting if we consider that for two sequences the Needleman-Wunsch algorithm generates an optimal alignment in $O(n^2)$ time \cite{Need}.) 

Even though it is unlikely that one can find a fast and accurate general algorithm for MSA, several heuristic approximation algorithms have been developed during last decades, because of the importance of this problem. One of the most frequently used among them is named Clustal and it is applying a progressive method: first, this algorithm is building a so-called "guide tree" to determine in which order it is the most practical to align the sequences, then using this tree, it is creating a multiple alignment. This process starts with aligning the two closest sequences optimally, and after that, in every step, a sequence that is not aligned yet will be aligned, or two sets of aligned sequences will be aligned to each other optimally \cite{Takacs}. Many other heuristic algorithms for MSA are applied widely which can use progressive methods like Clustal (e. g., T-Coffee), iterative methods (DIALIGN) or even tools (Hidden Markov models) of probability theory (POA) \cite{Cedric}.   

Summing up, it is known that MSA problem is hard in general, but it is still an interesting question that precisely when it begins to be really hard. One can assume that for length-1 (and perhaps even for length-2) sequences, it may not be that hard to find an optimal alignment. Furthermore, if an optimal alignment for short sequences can be determined in polynomial time, then it could also help to develop faster or more accurate heuristic algorithms. In this work, some new results regarding with aligning of short sequences are presented. In my opinion, it can be the most relevant result of this work that for length-1 sequences using arbitrary metric, as well as for length-2 sequences using some special metric, the optimum of the MSA problem can be easily, in the most trivial way determined. 

\pagebreak

\chapter*{2. Definitions and notations}
\addcontentsline{toc}{chapter}{2. Definitions and notations}

Let $\Sigma = \{a_1, \dots, a_n\}$ be a finite alphabet. A string over $\Sigma$ is called a \textit{sequence}.  $s_1'$ and $s_2'$ is an \textit{alignment} of sequences $s_1$ and $s_2$ if $(\forall \ i) \ s_i'$ is obtained from $s_i$ by inserting gaps (spaces, denoted by $-$) into or at either end of $s_i$ and by that, $s_1'$ and $s_2'$ have the same length. (It is assumed that $-$ is not an element of $\Sigma$.) Because of this definition, every character of $s_1'$ is uniquely corresponded to a character of $s_2'$.

Let \textit{l} be the common length of $s_1'$ and $s_2'$. The \textit{cost} of this alignment is $\displaystyle \sum_{i=1}^{l}\textit{d}(s_1'(i), s_2'(i))$, where \textit{d} is a \textit{score scheme} over $\Sigma \cup \{-\}$ and $s_j'(i)$ is the \textit{i}th character of $s_j'$. It is commonly required that a score scheme must satisfy triangle inequality: $\forall i, j, k: \ d(a_i, a_j) \leq d(a_i, a_k) + d(a_k, a_j)$. A frequently used score scheme is the so-called \textit{unit metric}, where $d(a_i, a_j) = 0$ if $i=j$ and 1 otherwise.  We call an alignment \textit{optimal} for two sequences if its cost is minimal among every possible alignments. 

The definiton of aligning two sequences can be easily generalized for more strings: let \textit{k} be the number of sequences to align. Let us insert gaps into or at either end of every strings so that they have the same \textit{l} length and in the proper order, write them under each other. We call this matrix size of \textit{k}x\textit{l} a \textit{multiple alignment} of these sequences. There are different \textit{scoring methods} how to define the \textit{cost} of a multiple alignment, perhaps the most often used one is the \textit{sum of pairs} method: using this, we get the required cost of an alignment as the sum of the costs of aligning the $\binom{k}{2}$ pairs from the aligned sequences. In a formula: if $s_1, \dots, s_k$ are sequences to align, then their sum of pair cost is $\displaystyle \sum_{i=1}^{k-1} \sum_{j=i+1}^{k} cost(s_i, s_j)$ \cite{Wang}.

\

\textit{Examples.} $i)$  Let $S$ be $S := \{CCG, GCG, CGC\}$. The following set of aligned sequences is a multiple alignment $\mathcal{A}$ of $S$:

\begin{center}
\begin{tabular}{c c c c}
C & C & G & $-$ \\
G & C & G & $-$ \\
$-$ & C & G & C \\
\end{tabular}
\end{center}

Using unit metric and considering the cost of the columns, cost($\mathcal{A}) = 3 + 0 + 0 + 2 = 5$.

\

$ii)$ Let $\Sigma$ now contain only two characters $(C$ and $G)$ with the following metric:

\

\begin{center}
\begin{tabular}{c|c|c|c}

$ $ & $C$ & $G$ & $-$ \\
\hline
$C$ & 0 & 2 & 1 \\
\hline
$G$ & 2 & 0 & 1 \\
\hline
$-$ & 1 & 1 & 0 \\
\end{tabular}
\end{center}

\

Let $S$ be $S := \{CG, GC, GG \}$. In this case, the following set is a multiple alignment $\mathcal{A}$ of $S$:

\begin{center}
\begin{tabular}{c c c}
$-$ & C & G \\
G & C & $-$ \\
G & $-$ & G  \\
\end{tabular}
\end{center}

Using the given metric, cost$(\mathcal{A})$ will be equal to $ 2 + 2 + 2 = 6$. 

\chapter*{3. Multiple sequence alignment for length-1 sequences}
\addcontentsline{toc}{chapter}{3. Multiple sequence alignment for length-1 sequences}

In this section, let us focus on aligning length-1 sequences (equivalently, characters of $\Sigma$). To prove a theorem regarding sequences of this kind, an important earlier result must be used:

\begin{thm}[\cite{Bon}] 
Let $U$ be a subset of a set $S$ of sequences over $\Sigma$ such that $U$ contains only identical sequences, and let $\mathcal{A}$ be an optimal alignment of $S$. Then $\displaystyle d(\mathcal{A}_U) = \sum_{u_i \in U} \sum_{\substack{u_j \in U \\ i < j}} d(u_i, u_j) = 0$ in $\mathcal{A}$.
\end{thm}

As an important corollary, this theorem is implying that it is enough to examine sets of sequences where these sequences are pairwise different, because in an optimal alignment, every instance of a given sequence is aligned identically.

The next definition will be used frequently throughout this work:

\begin{defi}
Let $S$ be a set of sequences that have the same length. $\mathcal{A}$ is the trivial alignment of $S$ if $\mathcal{A}$ is constructed by writing every sequnce under each other, without using gaps.
\end{defi} 

\section*{3.1. Multiple sequence alignment for length-1 sequences using unit metric}
\addcontentsline{toc}{section}{3.1. Multiple sequence alignment for length-1 sequences using unit metric}

\

The main result of this subsection is the next theorem:

\begin{thm} 
Using unit metric, there can not be a multiple sequence alignment for length-1 sequences that has less cost than their trivial alignment, and if we align $k$ different sequences, then the cost of an optimal alignment is $\binom{k}{2}$.
\end{thm}

\textit{Proof.}
Considering Theorem 1, it can be assumed that the characters that need to be aligned are pairwise different because if there were some identical ones among them, then all instances of a particular character would be aligned the same way.

It is easy to see that the trivial alignment of $k$ different characters has a cost of $\binom{k}{2}$: there are $\binom{k}{2}$ pairs among these characters and in every pair, there are two different sequences, so the cost of an aligned pair is always $1$.

If we assume that this alignment is not optimal, then the length of every aligned sequence must be at least $2$ in an optimal alignment, and we have to examine the cost of an alignment of this type. If this common length of aligned sequences is \textit{l} $\geq 2$, then the general structure of the \textit{n}x\textit{l} matrix of this multiple alignment is the next: $\forall 1 \leq$ \textit{i} $\leq$ \textit{l} there are $k_i$ characters in the \textit{i}th column ($\displaystyle \sum_{i=1}^{l} k_i = k$) and they are placed so that in every row, there is only one character and $l-1$ gaps (see Figure 1).

If we focus on the first column, we can establish that its cost is $\binom{k_1}{2} + (k-k_1)k_1$, since there are $k_1$ different characters with cost of $\binom{k_1}{2}$, and besides that, all of the $(k-k_1)$ gaps increases the cost by one with every alphabetical character. A similiar statement is true for every column, so the cost of this alignment: $\sum_{i=1}^{l} \binom{k_i}{2} + (k-k_i)k_i = k \sum_{i=1}^{l} k_i - \frac{\sum_{i=1}^{l} k_i}{2} - \frac{\sum_{i=1}^{l} k_i^2}{2} = k^2 - \frac{k}{2} - \frac{\sum_{i=1}^{l} k_i^2}{2}.$ Thus if we want to minimize the cost of this alignment then we have to maximize $\displaystyle \sum_{i=1}^{l} k_i^2$.

\

\begin{figure}[!h]
\begin{center}
\resizebox{6.0cm}{!}{
\begin{tabular}{c c c c}

$a_1$ & $-$ & $\dots$ & $-$ \\
$\dots$ & $\dots$ & $\dots$ & $\dots$ \\
$a_{k_1}$ & $-$ & $\dots$ & $-$ \\
$-$ & $a_{k_1 + 1}$ & $\dots$ & $-$ \\
$\dots$ & $\dots$ & $\dots$ & $\dots$ \\
$-$ & $a_{k_2}$ & $\dots$ & $-$ \\
$\dots$ & $\dots$ & $\dots$ & $\dots$ \\
$-$ & $-$ & $\dots$ & $a_{k_{l-1} + 1}$ \\
$\dots$ & $\dots$ & $\dots$ & $\dots$ \\
$-$ & $-$ & $\dots$ & $a_{k_l}$ \\
\end{tabular}}
\captionsetup{labelformat=empty}
\caption{Figure 1. A multiple alignment for length-1 sequences on $l$ columns.}
\end{center}
\end{figure}

\

By examining $\displaystyle (\sum_{i=1}^{l} k_i)^2$, we can see that it is equal to $k^2$, but at the same time, it is equal to $\displaystyle \sum_{i=1}^{l} k_i^2 + 2 \sum_{i=1}^{l} \sum_{\substack{j=2 \\ i < j}}^{l} k_i k_j$. From this, it is clear that $\displaystyle \sum_{i=1}^{l} k_i^2 \leq k^2$, and by that, the cost of this alignment can not be less than $\frac{k^2 - k}{2} = \binom{k}{2}$. $\square$

\textit{Note.} From the proof, it is also clear (by minimizing $\displaystyle \sum_{i=1}^{l} k_i^2$) that a multiple alignment for $k$ different length-1 sequences can not have a cost more than $k^2 - \frac{k}{2} - \frac{k^2}{2l}$ if the length of aligned sequences is $l$. Since $l \leq k$, the cost can be at most $k^2 - k$ and this limit can be reached indeed: if there is only one character in every column and in every row, then the cost will be $k(k-1) = k^2-k$.

\section*{3.2. Multiple sequence alignment for length-1 sequences using arbitrary metric}
\addcontentsline{toc}{section}{3.2. Multiple sequence alignment for length-1 sequences using arbitrary metric}

\

In this subsection, it will be shown that for length-1 sequences, we can use any metric as score scheme, the multiple sequence alignment problem still remains as easy as in case of unit metric.

\begin{thm}
Using arbitrary metric, there can not be a multiple sequence alignment for length-1 sequences that has less cost than their trivial alignment, and if we align $k$ different sequences, then the cost of an optimal alignment is equal to $C = \displaystyle \sum_{i=1}^{k} \sum_{\substack{j=2 \\ i<j}}^{k} d(a_i, a_j)$.
\end{thm}

\textit{Proof.}
Because of Theorem 1, it can be assumed again that every sequence has exactly one instance in the set $S$ of sequences to be aligned. If we consider the trivial alignment of the $S$, it is easy to see that its cost is equal to $C$. Induction for the number of the columns in a multiple sequence alignment will be used to show that any alignment can not have lower cost than $C$. 

Let be assumed that the trivial alignment is not optimal and let $\mathcal{A}$ denote an optimal alignment. Assuming that $\mathcal{A}$ is not the trivial alignment, $\mathcal{A}$ has $l$ columns where $l \geq 2$. It can be showed that $\mathcal{A}$ can not have exactly two columns, because in this case, trivial alignment would have a lower cost than $\mathcal{A}$ has.

Let be assumed to the contrary that $\mathcal{A}$ has exactly two columns; so there are $k_1$ sequences in the first column and $k_2$ in the second column, where $k_1 + k_2 = k$ and there is exactly one character in every row (see Figure 2).

\
\begin{figure}[!h]
\begin{center}
\resizebox{3.0cm}{!}{
\begin{tabular}{c c}
$a_1$ & $- $\\
$a_2$ & $-$ \\
$\dots$ & $\dots$ \\
$a_{k_1}$ & $-$ \\
$-$ & $a_{k_1 + 1}$ \\
$-$ & $a_{k_1 + 2}$ \\
$\dots$ & $\dots$ \\
$-$ & $a_{k}$ \\
\end{tabular}}
\captionsetup{labelformat=empty}
\caption{Figure 2. A multiple alignment for $k$ length-1 sequences on two columns.}
\end{center}
\end{figure}

\

It can be assumed without loss of generality that the sequences in the first column are $a_1, a_2, \dots, a_{k_1}$ and every other sequence are placed in the second row. If the cost of the first column of $\mathcal{A}$ is denoted by $cost(l_1)$, then \[cost(l_1) = \displaystyle \sum_{i=1}^{k_1} \sum_{\substack{j=2 \\ i<j}}^{k_1} d(a_i, a_j) + k_2 \sum_{i=1}^{k_1} d(a_i, -).\] \linebreak Similiarly, the cost of the second column is \[cost(l_2) = \displaystyle \sum_{i=k_1 + 1}^{k} \sum_{\substack{j=k_1 + 2 \\ i<j}}^{k} d(a_i, a_j) + k_1 \sum_{j=k_1+1}^{k} d(a_j, -),\] and $cost(\mathcal{A}) = cost(l_1) + cost(l_2)$. 

A lower bound for $cost(\mathcal{A})$ can be determined by pairing the $d(a_i, -)$ summands in $cost(l_1)$ to the summands of same form in $cost(l_2)$ and using triangle inequality. E. g., for a fix $i$ ($1 \leq i \leq k_1$) and $\forall j: \ k_1+1 \leq j \leq k$, it is true that $d(a_i, -) + d(a_j, -) \geq d(a_i, a_j)$, so $k_2 d(a_i, -) + \sum_{j=k_1 + 1}^{k} d(a_j, -) \geq \sum_{j=k_1 + 1}^{k} d(a_i, a_j)$. It is useful to notice that the summands on the right side of this inequality are exactly those ones that are not included in $cost(c_1)$ when we consider summands of the form of $d(a_i, a_j)$ for this fix $i$. 

By considering this inequality for every $1 \leq i \leq k_1$, the following lower bound can be given: \[k_2 \sum_{i=1}^{k_1} d(a_i, -) + k_1 \sum_{j=k_1 + 1}^{k} d(a_j, -) \geq \sum_{i=1}^{k_1} \sum_{j=k_1 + 1}^{k} d(a_i, a_j) \]

This is implying that $\displaystyle cost(\mathcal{A}) \geq \sum_{i=1}^{k_1} \sum_{\substack{j=2 \\ i<j}}^{k_1} d(a_i, a_j) + \sum_{i=k_1 + 1}^{k} \sum_{\substack{j=k_1 + 2 \\ i<j}}^{k} d(a_i, a_j) + \sum_{i=1}^{k_1} \sum_{j=k_1 + 1}^{k} d(a_i, a_j)  = \sum_{i=1}^{k} \sum_{\substack{j=2 \\ i<j}}^{k} d(a_i, a_j) = C$. It is assumed that the trivial alignment with cost $C$ is not optimal, therefore $\mathcal{A}$ can not be an optimal alignment of $S$. By this contradiction, it is proved that an optimal alignment of $S$ can not have exactly $2$ columns.

Using induction, let be assumed that it is shown $\forall i: 2 \leq i< l$ that an optimal alignment can not have exactly $i$ columns, and let $\mathcal{A}$ be an optimal alignment with $l$ columns. Considering the cost of the first two columns of $\mathcal{A}$, there are $k_1$ sequences in the first column and $k_2$ sequences in the second one. It is enough to prove that by merging these two columns, the cost of the new alignment is lower than the cost of $\mathcal{A}$. The cost of these columns in $\mathcal{A}$ is equal to $\displaystyle \sum_{i=1}^{k_1} \sum_{\substack{j=2 \\ i < j}}^{k_1} d(a_i, a_j) + (k - k_1) \sum_{i=1}^{k_1} d(a_i, -) + \sum_{i=k_1+1}^{k_2} \sum_{\substack{j=k_1+1 \\ i < j}}^{k_2} d(a_i, a_j) + (k - k_2) \sum_{i=k_1+1}^{k_2} d(a_i, -)$ (see Figure 3).

\

\begin{figure}[!h]
\begin{center}
\resizebox{3.0cm}{!}{
\begin{tabular}{c c}
$a_1$ & $- $\\
$a_2$ & $-$ \\
$\dots$ & $\dots$ \\
$a_{k_1}$ & $-$ \\
$-$ & $a_{k_1 + 1}$ \\
$-$ & $a_{k_1 + 2}$ \\
$\dots$ & $\dots$ \\
$-$ & $a_{k_1 + k_2}$ \\
$-$ & $-$ \\
$\dots$ & $\dots$ \\
$-$ & $-$ \\
\end{tabular}}
\captionsetup{labelformat=empty}
\caption{Figure 3. The first two columns of $\mathcal{A}$.}
\end{center}
\end{figure}

\

Let us focus on the first $k' = k_1 + k_2$ characters of these columns. It is an alignment of $\{a_1, a_2, \dots, a_{k'} \}$ on two columns and it was shown that if these sequences are aligned trivially instead of using two columns, then the cost of the alignment can not be higher. It means the following: 

\vspace{0.3cm}

$\displaystyle \sum_{i=1}^{k_1} \sum_{\substack{j=2 \\ i < j}}^{k_1} d(a_i, a_j) + k_2 \sum_{i=1}^{k_1} d(a_i, -) + \sum_{i=k_1+1}^{k'} \sum_{\substack{j=k_1+1 \\ i < j}}^{k'} d(a_i, a_j) \linebreak + k_1 \sum_{i=k_1+1}^{k'} d(a_i, -) + (k - k') \sum_{i=1}^{k_1} d(a_i, -) + (k - k') \sum_{i=k_1+1}^{k'} d(a_i, -) \geq \linebreak \sum_{i=1}^{k'} \sum_{\substack{j=2 \\ i<j}}^{k'} d(a_i, a_j) + (k - k')  \sum_{i=1}^{k'} d(a_i, -) $.

On the left side of this inequality, there is the cost of the first two columns of $\mathcal{A}$, while on the right side, there is the cost of the column that is constructed by merging the first two columns of $\mathcal{A}$. Therefore, a lower bound for $cost(\mathcal{A})$ is given by an alignment that has $l-1$ columns, implying that $\mathcal{A}$ can not be optimal. $\square$

\chapter*{4. Multiple sequence alignment for length-2 sequences}
\addcontentsline{toc}{chapter}{4. Multiple sequence alignment for length-2 sequences}

In this section, it will be shown that using unit metric, a set of length-2 sequences can not be aligned with less cost than their trivial alignment, however, this statement does not hold using arbitrary metric.

\begin{thm}
Using unit metric, there can not be a multiple sequence alignment for length-2 sequences that has less cost than their trivial alignment, and if we align $k$ different sequences $(s_1 = a_{i_1}a_{i_{k+1}}, s_2 = a_{i_2}a_{i_{k+2}}, \dots, s_k = a_{i_k}a_{i_{2k}})$, then the cost of an optimal alignment is $\displaystyle \sum_{j=1}^{k} \sum_{\substack{l=2 \\ j<l}}^{k} d(a_{i_j}, a_{i_l}) + \displaystyle  \sum_{j=k+1}^{2k} \sum_{\substack{l=k+2 \\ j<l}}^{2k} d(a_{i_j}, a_{i_l}) $.
\end{thm}

\textit{Proof.}
Let $S$ denote the set of sequences that need to be aligned. It is clear that the trivial alignment of $S$ has the cost written above, so this lower bound is accessible. In other words, it is enough to prove that for any $S$, a non-trivial alignment can not have less cost than the trivial one.

Let $\mathcal{A}$ be an alignment of $S$ on $t$ columns where $t \geq 3$. Let the rows of $\mathcal{A}$ be permuted so that those aligned sequences, where the indices of the two non-gap characters are the same, are placed under each other, forming a block of sequences (by this operation, the cost of $\mathcal{A}$ does not change). In every row of $\mathcal{A}$, there are exactly two characters and $t-2$ gaps, so there can be $\binom{t}{2}$ types of aligned sequences in $\mathcal{A}$, considering only the positions of the characters in a row. This implies that there will be $\binom{t}{2}$ (not necessarily non-empty) blocks after permuting the rows of $\mathcal{A}$. (E. g., if $t = 4$, then there are $\binom{4}{2} = 6$ blocks after row permuting, see Figure 4.) 

\

\begin{figure}[!h]
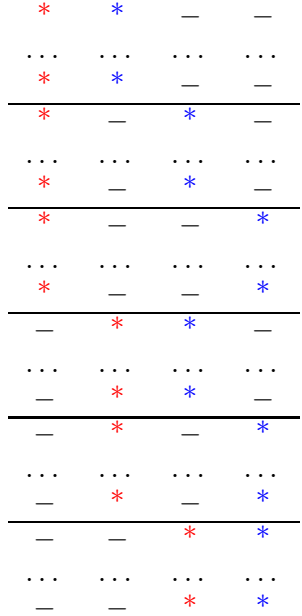

\begin{center}
\resizebox{4.0cm}{!}{
\begin{tabular}{c c c c}
\color{red} * & \color{blue} * & $-$ & $-$ \\
$\dots$ & $\dots$ & $\dots$ & $\dots$ \\
\color{red} * & \color{blue} * & $-$ & $-$ \\
\hline
\color{red} * & $-$ & \color{blue} *  & $-$ \\
$\dots$ & $\dots$ & $\dots$ & $\dots$ \\
\color{red} * & $-$ & \color{blue} *  & $-$ \\
\hline
\color{red} * & $-$ & $-$ & \color{blue} *  \\
$\dots$ & $\dots$ & $\dots$ & $\dots$ \\
\color{red} * & $-$ & $-$ & \color{blue} *  \\
\hline
$-$ & \color{red} * & \color{blue} *  & $-$ \\
$\dots$ & $\dots$ & $\dots$ & $\dots$ \\
$-$ & \color{red} * & \color{blue} *  & $-$ \\
\hline
$-$ & \color{red} * & $-$ & \color{blue} *  \\
$\dots$ & $\dots$ & $\dots$ & $\dots$ \\
$-$ & \color{red} * & $-$ & \color{blue} *  \\
\hline
$-$ & $-$ & \color{red} * & \color{blue} *  \\
$\dots$ & $\dots$ & $\dots$ & $\dots$ \\
$-$ & $-$ & \color{red} * & \color{blue} *  \\
\end{tabular}}
\captionsetup{labelformat=empty}
\caption{Figure 4. The structure of $\mathcal{A}$ after permuting its rows and making its block setting if $t = 4$. The red stars denote the sequences' first characters, while the blue ones denote their second letters. During the proof, an upper bound is given for the cost of aligning letters with the same color that are not aligned in $\mathcal{A}$ by using character-gap alignment costs that are included in cost$(\mathcal{A})$.}
\end{center}
\end{figure}

After making this block setting, it is clear that there are six types of aligned character pairs in $\mathcal{A}$:

$i)$ first characters of some sequences aligned with other sequences' first characters;

$ii)$ first characters of some sequences aligned with other sequences' second characters;

$iii)$ first characters of some sequences aligned with gaps;

$iv)$ second characters of some sequences aligned with other sequences' second characters;

$v)$ second characters of some sequences aligned with gaps;

$vi)$ gaps aligned with gaps.

\

In the trivial alignment $\mathcal{T}$, there are only pairs of type i) and iv), moreover, \textit{every} sequence's first character is aligned with each other in $\mathcal{T}$ (and it holds similiarly for \textit{every} second character of the sequences of $S$). Nevertheless, in a non-trivial alignment $\mathcal{A}$, there are aligned sequences whose first or second characters are not aligned with each other in $\mathcal{A}$. This implies that it is enough to give an upper bound for the cost of these characters in $\mathcal{T}$ that are aligned with each other in $\mathcal{T}$ but are not aligned with each other in $\mathcal{A}$, using parts of cost($\mathcal{A}$) for this bound (see Figure 5). (Because every part of cost($\mathcal{A}$) is nonnegative, if a bijection can be given between the letter-letter alignments in $\mathcal{T}$ that are not aligned in $\mathcal{A}$ and some other alignments of characters of $\mathcal{A}$ (not excluded character-gap alignments) so that the latter alignments have always at least as much cost as the former ones, then it means that cost$(\mathcal{A}) \geq$ cost($\mathcal{T}$).) 

\begin{figure}[!h]
\begin{center}
\resizebox{5.5cm}{!}{
\includegraphics{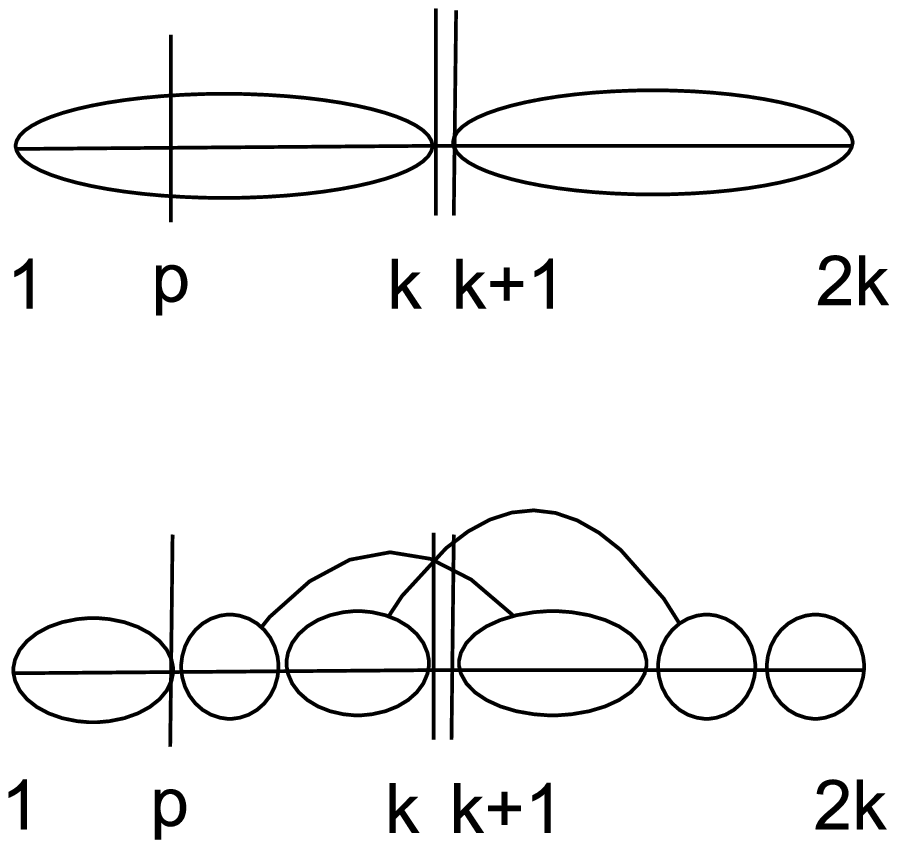}}
\captionsetup{labelformat=empty}
\caption{Figure 5. The general structure of character-character alignments of $\mathcal{T}$ (above) and $\mathcal{A}$, if $t = 4$. An ellipse is symbolizing that every pair of characters that are contained by the same ellipse are aligned with each other in the given alignment, moreover, in the figure of $\mathcal{A}$, two edge-connected ellipses are also aligned with each other. It can be seen that in $\mathcal{T}$, every pair of the first characters of sequences are aligned with each other and this statement is true for every pair of second characters of sequences, too. However, in $\mathcal{A}$, any pairs of the form $(a_{i_j}, a_{i_l})$ will not be aligned, where $1 \leq i_j \leq p$ and $p+1 \leq i_l \leq k$, implying that the cost of $d(a_{i_j}, a_{i_l})$, which is a part of cost($\mathcal{T}$) but not a part of cost($\mathcal{A}$), must be overestimated with a part of cost($\mathcal{A}$).}
\end{center}
\end{figure}

If $d$ denotes the unit metric, then the following inequality holds for every pair of sets $P, R$ on arbitrary alphabet (where $P$ and $R$ can contain a letter more than once): \[\displaystyle \sum_{a_{i_j} \in P} \sum_{a_{i_l} \in R} d(a_{i_j}, a_{i_l}) \leq |P|  \sum_{a_{i_l} \in R} d(a_{i_l}, -) = |P| |R|.\]

Using this inequality, a bijection mentioned above can be given: first, let be considered two sequences whose first characters ($a_i$ and $a_j)$ are not aligned in $\mathcal{A}$. (It can be assumed that $a_j$ has bigger column index.) This implies that the element that is in the intersection of the row of $a_j$ and the column of $a_i$ must be a gap. $d(a_i, a_j) \leq d(a_i, -)$, so the cost of the alignment of $a_i$ and $a_j$ in $\mathcal{T}$ can be estimated by the cost of the alignment of two characters in $\mathcal{A}$.

Similiarly, if two sequences are considered whose second characters  $(a_i$ and $a_j)$ are not aligned in $\mathcal{A}$, then (assuming that $a_j$ has bigger column index) the element in the intersection of the row of $a_i$ and the column of $a_j$ must be a gap. The same estimation can be given like before, meaning that the cost of the alignment of $a_i$ and $a_j$ in $\mathcal{T}$ is less or equal to the cost of a character-gap alignment in $\mathcal{A}$.

Considering the block setting of $\mathcal{A}$, let $B_i$ and $B_j$ two blocks whose sequences' first characters are not aligned in $\mathcal{A}$. Assuming that the first characters of sequences in $B_j$ have bigger column index, there must be $|B_j|$ gaps in the intersection of the column of the first characters of sequences in $B_i$ and the rows of $B_j$. If we denote the first letters of the sequences of $B_i \ (B_j)$ by $a_{b_i} \ (a_ {b_j})$, then (because of the statements of the latter two paragraphs) the following holds: \[\sum_{b_i \in B_i} \sum_{b_j \in B_j} d(a_{b_i}, a_{b_j}) \leq |B_j| \sum_{b_i \in B_i} d(a_{b_i}, -) = |B_i| |B_j| \]

\begin{figure}[!h]
\begin{center}
\resizebox{2.5cm}{!}{
\begin{tabular}{|c|c|c|c|}
\hline
$1$ & $2$ & $-$ & $-$ \\
\hline
$1$ & $-$ & $2$ & $-$ \\
\hline
$1$ & $-$ & $-$ & $2$ \\
\hline
$-$ & $1$ & $2$ & $-$ \\
\hline
$-$ & $1$ & $-$ & $2$ \\
\hline
$-$ & $-$ & $1$ & $2$ \\
\hline
\end{tabular}}
\captionsetup{labelformat=empty}
\caption{Figure 6. The block setting of $\mathcal{A}$ if $t=4$, denoting only that an element is the first/second character of its aligned sequence or a gap. E.g., the first element of the first row in the block setting and the second element of the fourth row (which are denoting the first characters of some sequences) are not aligned in $\mathcal{A}$, so the cost of their alignment with each other, which is a part of cost$(\mathcal{T})$ but not a part of cost$(\mathcal{A})$, must be overestimated with a part of cost$(\mathcal{A})$, namely, with the cost of aligning the block setting's first element of the first row with the gaps in the first element of the fourth row.}
\end{center}
\end{figure}

Besides that, a similiar result can be established if we consider two blocks whose sequences' second characters are not aligned, using the gaps of the block that has the column with smaller column index (see Figure 6). By these estimations, it is clear that this assignment between the character-character alignments in $\mathcal{T}$ that are not present in $\mathcal{A}$ and character-gap alignments in $\mathcal{A}$ eventuates that the latter costs in $\mathcal{A}$ can not be less than the corresponding costs in $\mathcal{T}$. It also must be examined that this assignment is a bijection, i. e. there are no character-gap alignments that are used multiple times.

A set of gaps in the block setting are considered in an estimation if and only if some characters in the block that is containing these gaps and some characters from another block that are aligned in the same column must be aligned in $\mathcal{T}$ but they are not aligned in $\mathcal{A}$. This is implying that these gaps are not used in estimations like above more times than the alignment of this gap set with the rest of the given column, therefore the former assignment is a bijection, implying that cost$(\mathcal{A}) \geq$ cost$(\mathcal{T})$. $\square$

\

\textit{Note.} It is worthy of note that during the proof, the following special property of unit metric has been used only: $\forall a_i, a_j \in \Sigma: d(a_i, a_j) \leq d(a_i, -)$. It follows that there can not be an alignment for a set of length-2 sequences that has less cost than their trivial alignment, if a metric that has the same property is being used. 

\

As the next example shows, trivial alignment will not always be optimal for length-2 sequences if an arbitrary metric can be used. Let $\Sigma$ contain two characters $(C$ and $G)$ with the same metric on $\Sigma$ as in the Example $ii)$ at the end of Section 2, moreover, let $S$ be also the same: $S = \{CG, GC, GG \}$. The trivial alignment of $S$ has a cost of $8$, but as it has been shown, there is an alignment of $S$ that has only cost of $6$ (see Figure 7).

\

\begin{figure}[!h]
\begin{center}
\resizebox{1.5cm}{!}{
\begin{tabular}{c c}
C & G \\
G & C \\
G & G \\
\end{tabular}}
\quad
\resizebox{2.0cm}{!}{
\begin{tabular}{c c c}
$-$ & C & G \\
G & C & $-$ \\
G & $-$ & G  \\
\end{tabular}}
\captionsetup{labelformat=empty}
\caption{Figure 7. The trivial and an optimal alignment of S.}
\end{center}
\end{figure}

\textit{Note.} In Section 2, it was shown that we can easily determine the minimum cost of a set to be aligned if it includes only length-1 sequences, moreover, we also can construct an optimal alignment in the most trivial way using any metric. We have also seen that for length-2 sequences, the trivial alignment is optimal if unit metric is used but it is not optimal for arbitrary metric. Besides that, it is also known that trivial alignment is not always optimal for length-3 sequences even using unit metric.

As in Example $i)$ at the end of Section 2, let $S$ be the follow: $S = \{CCG, GCG, \linebreak CGC\}$. Using unit metric, the cost of the trivial alignment is $6$, but it is not optimal: as we have seen, there is a non-trivial alignment $\mathcal{A}$ of $S$ so that cost$(\mathcal{A})$ is only $5$ (see Figure 8).

\

\begin{figure}[!h]
\begin{center}
\resizebox{2.0cm}{!}{
\begin{tabular}{c c c}
C & C & G \\
G & C & G \\
C & G & C \\
\end{tabular}}
\quad
\resizebox{2.6cm}{!}{
\begin{tabular}{c c c c}
C & C & G & $-$ \\
G & C & G & $-$ \\
$-$ & C & G & C \\
\end{tabular}}
\captionsetup{labelformat=empty}
\caption{Figure 8. The trivial and an optimal alignment of S.}
\end{center}
\end{figure}

\chapter*{5. Open questions}
\addcontentsline{toc}{chapter}{5. Open questions}

In this work, it was shown that the multiple sequence alignment problem is "easy" for length-1 sequences and also for length-2 sequences in special cases. Since we know that the general problem is $\mathbf{NP}$-complete, it is still an interesting question that for how long sequences MSA starts to become a real hard problem? It is probably another open problem that in case of length-2 sequences, how can those metrics be characterized for which trivial alignment is always optimal for arbitrary alphabet? 

\chapter*{Acknowledgement}

I would like to return thanks Dr. Vince Grolmusz for his many suggestions and much help in order to make the quality of this work as high as it was possible. I am completely sure about that the current version of this work could not be established without his proposals and inspirations.


\renewcommand{\refname}{References} 

\bibliographystyle{plain}

\bibliography{tgf} 


\end{document}